# Quantification of $^{242}$Pu with a Microcalorimeter Gamma Spectrometer

David J. Mercer[a,*], Ryan Winkler[a], Katrina E. Koehler[a,b], Daniel T. Becker[c], Douglas A. Bennett[d], Matthew H. Carpenter[a], Mark P. Croce[a], Krystel Iris M. de Castro[a,e], Eric A. Feissle[a], Joseph W. Fowler[c,d], Johnathon D. Gard[c], John A. B. Mates[d], Daniel G. McNeel[a], Nathan J. Ortiz[c,d], Daniel Schmidt[d], Katherine A. Schreiber[a], Daniel S. Swetz[d], Joel N. Ullom[c,d], Leila R. Vale[d], Sophie L. Weidenbenner[a], and Abigail L. Wessels[c]

[a]Los Alamos National Laboratory, Los Alamos, New Mexico
[b]Houghton College, Houghton, New York
[c]University of Colorado, Boulder, Colorado
[d]National Institute of Standards and Technology, Boulder, Colorado
[e]Ateneo de Manila University, Quezon City, Philippines
*Corresponding author, Mercer@LANL.gov

**Abstract**: We report measurements of the 103-keV and 159-keV gamma ray signatures of $^{242}$Pu using microcalorimetry. This is the first observation of these gamma rays in a non-destructive measurement of an unprepared sample, and so represents an important advance in nuclear material accountancy. The measurement campaign also serves as the first demonstration of a field campaign with a portable microcalorimeter gamma-ray spectrometer. For the 103-keV gamma ray we report an improved centroid energy and emission probability.

## Introduction

Plutonium-242 is a challenge for nondestructive assay (NDA) because of its low specific activity and low gamma emission probability per decay. Neutron multiplicity methods are suitable for quantification but can only be used after the isotopic ratios have been determined [1]. For these ratios, one must depend on destructive measurements, correlation estimates (which can be inaccurate for high-burnup samples) [2], or direct measurement of the gamma ray signatures. The latter has never been achieved successfully except with carefully prepared, thin, unshielded laboratory samples. The only direct gamma rays from $^{242}$Pu are at 45, 103, and 159 keV, and all are weak. The lowest energy line is of limited use because of potential attenuation within the sample and/or container. The two higher-energy lines are masked by gamma rays from $^{238-241}$Pu, $^{241}$Am, and/or fluorescent X-rays, making them all but impossible to measure with high-purity germanium (HPGe) detectors except under very special circumstances.

Microcalorimetry (μCal) [3][4][5][6] offers a potential method for NDA of $^{242}$Pu. The energy resolution of μCal detectors is extraordinary, with 70-eV full-width at half-maximum (FWHM) now routinely achieved. Although efficiency is lower, the resolution allows for direct observation of the 103- and 159-keV signatures. In this demonstration we use these signatures to quantify $^{242}$Pu in a packaged item with no special sample preparation.

## Prior Gamma Measurements of $^{242}$Pu

In 1986, Vaninbroukx et al. [7] measured the γ emission probabilities for $^{242}$Pu using HPGe detectors with a FWHM resolution of 495–520 eV at 122 keV, which is near the best achievable for HPGe. Specially-prepared ultra-pure samples (99.85% Wt% $^{242}$Pu) were measured in a configuration designed for negligible photon attenuation. The $^{242}$Pu peaks were not resolvable from interfering components from $^{241}$Am, $^{241}$Pu, and $^{240}$Pu decay, whose contributions were computed and subtracted.

In 2011, Berlizov et al. performed measurements on a 99.7 Wt% $^{242}$Pu sample with a HPGe detector [8] in response to a suspected discrepancy. Their value for the $\gamma_{159}$ emission probability was 35% smaller than Vaninbroukx et al.'s. No attempt was made to use the 45- and 103-keV lines. They considered the 159-keV line as "the only practical alternative" for quantitative analysis of $^{242}$Pu, but no practical method was developed.

In 2012, Wang made measurements on a 99.97% Wt% $^{242}$Pu sample evaporated onto a thin foil [9]. An HPGe detector was used in coincidence with a Si(Sb) alpha detector; the α-γ coincidence mode allowed a reduction of interferences, mostly from $^{241}$Pu β-decay. Values for the emission probabilities were determined for all three $^{242}$Pu lines. The results are shown in Table I in comparison with Vaninbroukx et al., Berlizov et al., and the present work.

Table I: Reported γ-emission probabilities for $^{242}$Pu lines

| Reference | 45 keV | 103 keV | 159 keV |
|---|---|---|---|
| Vaninbroukx 1986 | 3.72(7)E-4 | 2.63(9)E-5 | 2.98(20)E-6 |
| Berlizov 2011 | — | — | 2.20(8)E-6 |
| Wang 2012 | 4.37(6)E-4 | 2.79(8)E-5 | 2.25(8)E-6 |
| Present Work* | | 2.69(3)E-5 | 2.10(14)E-6 |

*Explained in Quantification Results section

In 2016, Bates et al. successfully observed the 45-keV line with a metallic magnetic calorimeter (MMC, a type of microcalorimeter) [10]. The sample was 10.81 Wt% $^{242}$Pu, prepared as a solution dried onto a thin foil, and measured through a low-attenuation window. A FWHM resolution of 140 eV was achieved, and a quantitative measurement of the $^{242}$Pu concentration was reported in agreement with the declared value. This was a remarkable achievement, but the sample preparation and low-attenuation requirements are impractical for most NDA applications.

### Relocation and Installation of the Instrument

For our measurements of $^{242}$Pu, we used the micro-calorimeter array instrument SOFIA (Spectrometer Optimized for Facility Integrated Applications) [11]. The instrument was designed to be re-locatable and this was our first experience with moving it to a different site. The instrument was moved from Los Alamos National Laboratory (LANL) Technical Area 35 Building 02 to LANL Technical Area 55 Building PF-4. Figure 1 is a photograph of the instrument and supporting hardware loaded for transport. Major components consist of (left to right in Figure 1) a compact milliKelvin cryostat housing the detector, electronics rack, and helium compressor.

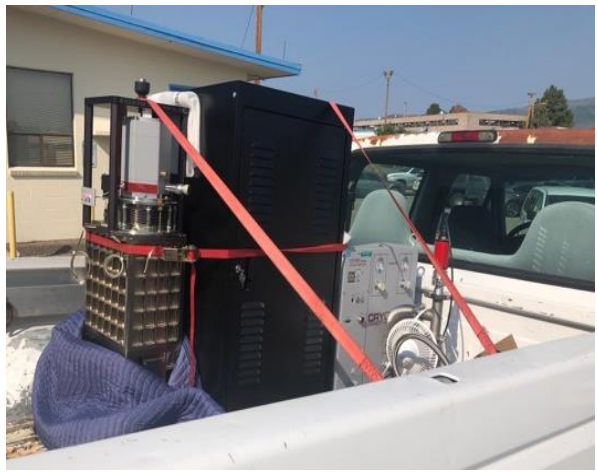

Figure 1. SOFIA and supporting hardware loaded for transport.

SOPHIA uses no liquid cryogens and requires only 220V, single phase power for the air-cooled helium pulse tube cryocooler. Suitable electrical power was readily available in the new location. Existing work authorization for gamma ray measurements in PF-4 was determined to cover operations as the instrument presents no unique hazards. No significant issues were encountered during the move and installation, and performance was verified to be consistent with that observed in the TA-35 laboratory.

### $^{242}$Pu Sample Characteristics

The $^{242}$Pu item used for the present measurements is "STDB242C8," consisting of 113.6 g of PuO$_2$ (99.75 g of Pu) in a steel "food-pack" can of unspecified wall thickness inside a SAVY container [12]. At the time of measurement the item was 86.85 Wt% $^{242}$Pu (see Table II). The packaged item was simply placed in front of the instrument with no preparation, consistent with a routine NDA procedure, as shown in Figure 2. It is notable that all prior gamma measurements of $^{242}$Pu involved bare or very lightly attenuated samples, and most involved ultra-pure (>99.5 Wt%) $^{242}$Pu, but the purpose of these measurements was to demonstrate the utility of the μCal instrument for routine nondestructive quantification.

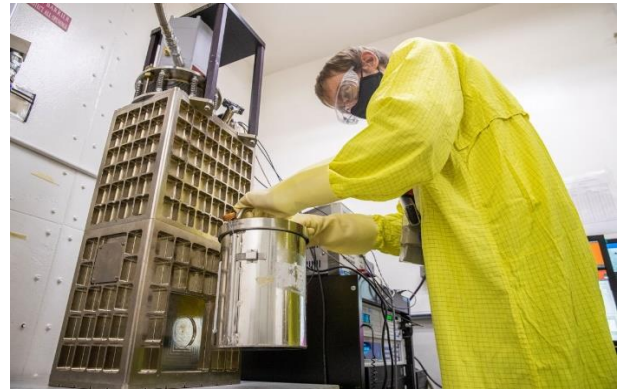

Figure 2. Sample STDB242C8 being placed in its measurement configuration by co-author Eric Feissle.

Table II: Declared isotopic concentrations for STDB242C8.

| Isotope (units) | 10-Nov-88* | 30-Aug-21 |
|---|---|---|
| $^{238}$Pu (Wt%) | 0.996 ± 0.007 | 0.790 ± 0.006 |
| $^{239}$Pu (Wt%) | 1.437 ± 0.008 | 1.476 ± 0.008 |
| $^{240}$Pu (Wt%) | 9.97 ± 0.02 | 10.21 ± 0.02 |
| $^{241}$Pu (Wt%) | 3.10 ± 0.01 | 0.652 ± 0.002 |
| $^{242}$Pu (Wt%) | 84.48 ± 0.02 | 86.85 ± 0.02 |
| $^{244}$Pu (Wt%) | 0.020 ± 0.001 | 0.021 ± 0.001 |
| $^{241}$Am (μg/gPu) | 350 ± 0.8 | 24793 ± 57 |

*Values were measured by mass spectrometry on this date [13] and are decay-corrected to match the μCal measurement dates.

### 103-keV Peak

The spectrum in this region is complex with overlapping features, but the energy resolution (70-eV FWHM) achieved with SOPHIA is sufficient to extract the area of the $^{242}$Pu peak. Figure 3 shows a detail of this region with the peak highlighted. This is the clearest observation of this peak from $^{242}$Pu decay ever reported, and this the first time it has been observed without special sample preparation. Vaninbroukx et al. and Berlizov et al. observed an unresolved multiplet including $^{241}$Am, $^{240}$Pu, $^{241}$Pu, and $^{242}$Pu components, Wang likely observed the peak via α-γ coincidence measurements but did not specifically report it, and Bates et al. did not report observations in this region likely due to limited efficiency.

We find that the $^{242}$Pu peak centroid is at 103.436 ± 0.003 keV. This is more than 1 sigma lower in energy than the 103.499 ± 0.035 keV reported by Schmorak et al. [14], who discuss "impurities which made it more difficult to establish the peak centroids," so the difference is unsurprising. The actual centroid position is fortunate for quantification because the separation from interfering peaks is increased from what was expected.

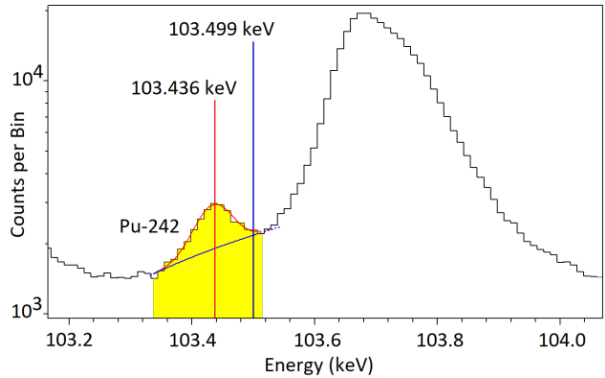

Figure 3: Spectrum from the first run detailing the 103-keV region. The $^{242}$Pu peak of interest is highlighted. Note that this peak appears at a lower energy than reported by Schmorak et al.

Although SOFIA is able to measure relative centroid energies with a precision better than 0.0005 keV, a limiting factor is the uncertainty in adopted centroids of nearby peaks needed for calibration, especially $^{241}$Am. Our $^{242}$Pu centroid shown above is interpolated using $^{241}$Am at 102.966 ± 0.005 keV [15] and Pu Kα1 at 103.734 ± 0.0006 keV [16] Many databases adopt the $^{241}$Am centroid as 102.980 ± 0.020 keV [17]; if this value is used instead in our interpolation, then our calculated $^{242}$Pu centroid becomes 103.440 ± 0.008 keV. A secondary limitation for centroid determination is nonlinearities that are potentially introduced during processing of the μCal data (see Yoho et al. 2020 [18]), which is a topic of future interest if this instrument will be used to improve nuclear data tables. Even with these limitations, our new value is considerably more reliable than the previously adopted value.

We extracted the peak areas in this region using the software SAPPY [5][19][20][21]. Component fits are shown in Figure 4. The 103.436-keV peak and nearby cluster are fit very well. The high-energy side of the $^{241}$Am 102.966-keV peak is fit poorly, suggesting that unexplored structure(s) are likely present. This is a topic for future investigation but has a negligible effect on the $^{242}$Pu peak analysis.

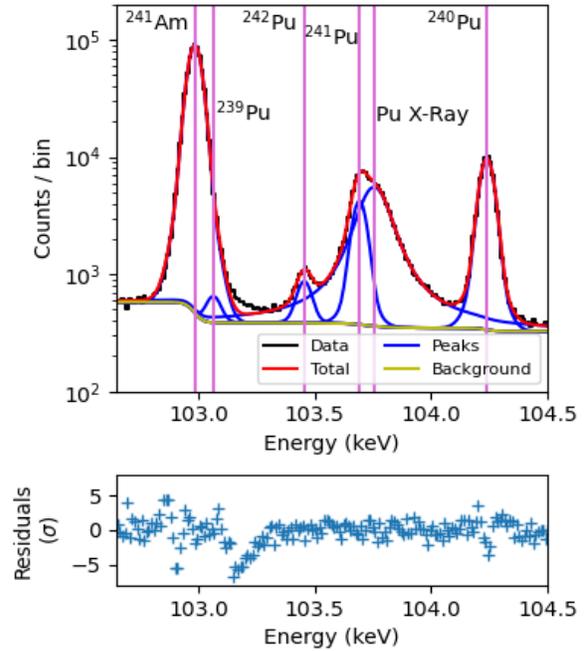

Figure 4: Fits of the 103-keV region using the software SAPPY.

The nearby $^{241}$Pu 103.68-keV peak is of practical interest for plutonium isotopic analysis. It arises directly from $^{241}$Pu α-decay ($^{241}$Pu → $^{237}$U), unlike most signatures for this radionuclide, which involve the $^{241}$Pu → $^{241}$Am → $^{237}$Np and/or $^{241}$Pu → $^{237}$U → $^{237}$Np chains. The direct decay means that the signature is time-independent and therefore reliable for very freshly separated plutonium. Fluorescent Pu Kα1 X-rays at 103.74 keV interfere with this $^{241}$Pu peak (as well as the $^{242}$Pu peak), but SAPPY is able to isolate the individual components.

The $^{240}$Pu peak at 104.23 keV is also of practical interest for improving precision of isotopic analysis. This peak is cleanly separated and its area is easy to extract. A drawback is interference from Sn kα1 and kα2 escape lines at 104.03 and 104.25 keV (too weak to be seen here but stronger for lower-burnup samples). Software techniques to deconvolve the escape peaks will be necessary to take full advantage of this $^{240}$Pu signature.

### 159-keV Peak

The 159-keV region includes cleanly-separated peaks from $^{242}$Pu at 159.02 keV, $^{241}$Pu at 159.96 keV (which arises directly from $^{241}$Pu α-decay), and $^{240}$Pu at 160.31 keV. See Figure 5. Vaninbroukx et al., Berlizov et al., and Wang observed this region as an unresolved triplet, and Bates did not report observations in this region. Our measurement of the centroid energy is $159.032 \pm 0.042$ keV, which is consistent with Berlizov's value of $159.018 \pm 0.016$ keV [8] but our measurement does not improve upon their uncertainty. As with the 103-keV peak, our uncertainty is limited by systematic uncertainty in the adopted values for nearby peaks needed for calibration. The FWHM energy resolution is 76 eV in this region. Because the peaks are well-separated, no special technique is required for area extraction.

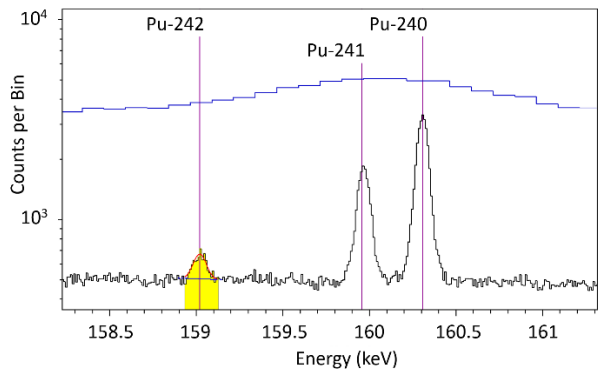

Figure 5: Spectrum detailing the 159-keV region. The peak of interest is highlighted. A typical-quality HPGe spectrum of the same region and same item is shown above the μCal spectrum.

### 45-keV Peak

We were unable to observe the 44.91-keV peak from $^{242}$Pu. Although our energy resolution is 60 eV at this energy, attenuation from the thick PuO$_2$ sample and thick steel container reduced the efficiency of detection to a level far below what would be necessary to see the peak. A $^{152}$Eu source, which produces Gd and Sm X-rays, was added during Run 4 to confirm the energy calibration and resolution in this region.

### Efficiency Calibration

Relative efficiency calibration as a function of energy is necessary for accurate quantification even though the extrapolation distance from $^{240}$Pu (or $^{241}$Pu) to $^{242}$Pu peak energies is less than 1.5 keV. The calibration involves the intrinsic detector efficiency, attenuation by the steel container, and self-attenuation within the PuO$_2$ sample. Intrinsic detector efficiency is determined using a separate $^{133}$Ba source and includes four free parameters in the form $\exp(a_0 + a_1 \ln E + a_2 \ln^2 E + a_3 \ln^3 E)$ where $E$ is γ-ray energy in keV. Two additional free parameters that account for container thickness and sample self-attenuation are found using a single gamma ray peak from $^{237}$U, two each from $^{238}$Pu and $^{240}$Pu, four from $^{241}$Pu, and seven from $^{241}$Am, using the test item itself as the source of radiation. The calibration assumes that the declared values in Table II are correct (excepting $^{242}$Pu, which is treated as an unknown). Many other peaks are available, but for simplicity we avoided any that involve multiple decay channels (such as 208.00 keV, which has components from both $^{237}$U and $^{241}$Am). Net peak areas were extracted using SAPPY in the complex 103-keV region. A simple tailless Voigt function is used for isolated singlet peaks. The result for Run 1 is shown in Figure 6. The units on the efficiency axis are detected counts per photon emitted from the radionuclide.

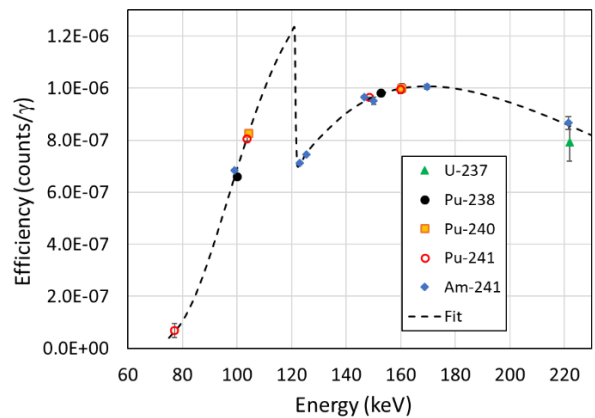

Figure 6: Efficiency calibration curve derived from five radionuclides within the sample. Some points are overlapping. The discontinuity near 120 keV is from the plutonium K-edge.

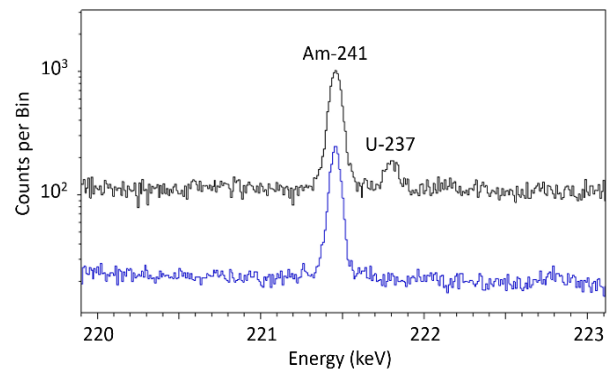

Figure 7: The 221.80-keV peak from $^{237}$U is cleanly separated from the 221.46-keV $^{241}$Am peak. A spectrum from a pure americium source is shown below, which lacks the $^{237}$U peak.

The single peak from $^{237}$U β-decay at 221.80 keV, shown in Figure 7, is of little value for the efficiency calibration but is of academic interest because it has never before been observed in a direct decay without overlap from the adjacent $^{241}$Am peak. The observation is a testament to the excellent resolution of μCal (80-eV FWHM at this energy). This peak arises from the 281.37-keV level in $^{237}$Np, which is apparently not populated by α-decay of $^{241}$Am.

## Quanitfication Results

The sample measurement was repeated four times. Run duration varied between 12.5 h and 21.4 h. Because the efficiency calibration relies on declared Wt% values for $^{237}$U, $^{238}$Pu, $^{240}$Pu, $^{241}$Pu, and $^{241}$Am, it is most meaningful to report $^{242}$Pu quantity relative to one of these. $^{240}$Pu is the second-most abundant radionuclide in the sample (10.21 Wt%) and it produces well-separated high-statistics peaks very close to 103 and 159 keV, so it will serve as the denominator. Results are shown in Table III. The $\gamma_{103}$ and $\gamma_{159}$ emission probabilities from Wang (Table I) are used in the calculations. The uncertainties reported in the table are from our measurements only; they do not include Wang's uncertainties. Our uncertainties are 1σ, and are dominated by counting statistics for the $^{242}$Pu peaks, but also include statistical uncertainty from the subtracted background and efficiency curve.

Table III: Quantification Results.

| Measurement | $^{242}$Pu/$^{240}$Pu Weight Ratio | |
| --- | --- | --- |
| | from 103 keV | from 159 keV |
| Run 1 | 8.15 ± 0.16 | 8.47 ± 1.16 |
| Run 2 | 8.39 ± 0.12 | 6.77 ± 0.74 |
| Run 3 | 8.21 ± 0.19 | 10.37 ± 1.48 |
| Run 4 | 8.31 ± 0.14 | 8.21 ± 0.97 |
| Weighted Avg. | 8.29 ± 0.07 | 8.30 ± 0.50 |

For both energies the weighted average results are slightly lower than the declared $^{242}$Pu/$^{240}$Pu weight ratio of 8.50 ± 0.02 (derived from Table II). The most precise results are achieved using the 103-keV peak despite the need to deconvolve the overlapping spectral region. Berlizov et al. [8] asserted that this peak is intractable with HPGe detectors; however, with μCal's excellent resolution, not only is use of the 103-keV peak possible, it is likely the best option for quanitification in many circumstances.

We were able to extract the 103-keV net area with a 1σ statistical uncertainty of 2% after a 12.5-hour measurement with a 100-gram sample. Better sensitivity may be possible with samples that are more diffuse, such as solutions, reducing interference from the Pu X-ray at 103.74 keV. Interference from the $^{241}$Pu peak at 103.68 keV also limits sensitivity. Due to these interferences, and barring substantial technological advances, the 103-keV $^{242}$Pu peak will be useful only for items with high Wt% $^{242}$Pu; the peak would be challenging to extract for weapons-grade and even high-burnup reactor-grade Pu with SOPHIA's current capabilities. The 159 keV region requires no deconvolution and suffers no significant interferences, but the $\gamma_{159}$ emission probability is less favorable for quantification. We were able to extract the 159-keV $^{242}$Pu peak with a 1σ uncertainty of 14% after a 12.5-hour measurement. This higher-energy peak may have an advantage for quantification of high-mass or heavily-shielded items, and a planned upgrade to double SOPHIA's efficiency will make its use more practical. For very thin containment, the 44.91-keV peak from $^{242}$Pu may become available for quantification, as demonstrated by Bates et al.[10], but a steel thickness of just 1/8 inch attenuates this peak by a factor of 1000, so it will remain unavailable in many NDA scenarios.

## Conclusion

We have successfully demonstrated the utility of μCal for $^{242}$Pu isotopic quantification, which meets a long-standing NDA challenge. This is the first time the 103- and 159-keV gamma ray peaks have been observed in a sample that was not specially prepared, and is the highest-resolution observation ever reported by any method. We have also found a more precise value for the 103-keV centroid, a more precise value for the $\gamma_{103}$ emission probability, and demonstrated portability of the SOFIA instrument.

Our 103-keV results are most consistent with a $\gamma_{103}$ emission probability of 2.69 (3) × 10$^{-5}$, which is between the values of Vaninbroukx et al. [7] and Wang [9] and within 1σ of both of them. Likewise, our 159-keV results are most consistent with a $\gamma_{159}$ emission probability of 2.10 (14) × 10$^{-6}$, which is within 1σ of both Wang and Berlizov et al., but disagrees with Vaninbroukx et al. We improve upon the uncertainty for the emission probability for $\gamma_{103}$ but not for $\gamma_{159}$. See Table I.

## Acknowledgments

This work was supported by Technology Evaluation & Demonstration funding from Los Alamos National Laboratory (LANL). The SOFIA instrument was developed under the U. S. Department of Energy, Office of Nuclear Energy, Material Protection Accounting and Control Technologies (MPACT) Program. LANL is managed by Triad National Security, LLC under Contract No. 89233218CNA000001 with the U.S. Department of Energy/ National Nuclear Security Administration. This work would not be possible without the partnership of the University of Colorado and the National Institute of Standards and Technology (NIST) Quantum Sensors Group, and support from the NIST Innovations in Measurement Science program and the DOE NEUP program. The U.S. Government retains a non-exclusive, irrevocable, worldwide license to publish or reproduce this manuscript or to allow others to do so.